\documentstyle[12pt]{article}

\def\be{\begin{equation}}
\def\ee{\end{equation}}
\def\ba{\begin{array}{c}}
\def\ea{\end{array}}
\def\p{\partial}
\def\ben{\[ }
\def\een{\] }

\begin{document}

\titlepage
.

\vspace*{4cm}

\begin{center}


{\Large \bf Solvability and ${\cal PT}-$symmetry in a
 double-well model with
point interactions
 }

\vspace{1.8truecm}

{\bf Miloslav Znojil} and
 {\bf V\'{\i}t Jakubsk\'{y}}

\vspace{1.3truecm}

 Department of Theoretical Physics,
  Nuclear Physics Institute,
Academy of Sciences,

 250 68
\v{R}e\v{z}, Czech Republic

\vspace{1.truecm}

e-mails: znojil@ujf.cas.cz and jakub@ujf.cas.cz

\end{center}



\vspace{0.5truecm}
\newpage
.

\vspace{4.truecm}

\section*{Abstract}

The concept of point interactions offers one of the most suitable
guides towards a quantitative analysis of properties of certain
specific non-Hermitian (so called ${\cal PT}-$symmetric)
quantum-mechanical systems. This is illustrated on a double-well
model, an easy solvability of which is shown to lead to a clear
picture of the mechanisms of the unavoided level crossing and of
the spontaneous ${\cal PT}$ symmetry breaking at a certain
strong-non-Hermiticity boundary. Below this limit the model is
shown suitable for an explicit illustration of technicalities
related to the standard probabilistic physical interpretation of
bound states in ${\cal PT}-$symmetric quantum mechanics.

 \vspace{0.5truecm}

PACS 03.65.Fd   03.65.Ca  03.65.Ge 03.65.Bz

\newpage
.

\vspace{1.truecm}

\section{Introduction}

A consistent probabilistic interpretation of bound states
$|\psi\rangle$ in quantum mechanics requires that their norm is
conserved in time \cite{Messiah}. In the light of the well known
Stone's theorem this means that their time-evolution must be
unitary so that the generator of this evolution (= the physical
Hamiltonian $H_{p}$) cannot be, apparently, non-Hermitian. Still,
the concept of a ''non-Hermitian Hamiltonian" $H_{n}\neq
H_{n}^\dagger$ is in a more or less current use in many
mathematical considerations (e.g., in perturbative calculations
\cite{BW} or in the schematic models of gravity \cite{grav} or
supersymmetry \cite{susy}) as well as in the various
phenomenological studies (e.g., in  nuclear physics \cite{Geyer},
solid-state physics \cite{HN}, particle physics and field theory
\cite{ Kleefeld} or in the model-building in magnetohydrodynamics
\cite{MHD} and cosmology \cite{ap}).

In such a context the use of the solvable point interactions seems
to offer a natural background for a simplification of some
mathematical challenges \cite{Kurasov} or of technical problems
\cite{Fei} as well as for an improvement of the feasibility of
numerical calculations \cite{DW} and for a deepening of the
related physical interpretation of the underlying dynamical
processes \cite{Weigert,BBJ}. In our paper we shall address
several aspects of non-Hermiticity in this framework, especially
those which were recently studied using the comparatively less
easily tractable piecewise constant potentials \cite{sqw}. We feel
encouraged by the success and impact of these studies which
already found applications in supersymmetric quantum mechanics
\cite{sqwapp} and which seem to exhibit certain promising
phenomenological (e.g., localization) properties \cite{sqwmetric}.

We are persuaded that the class of the point-interaction models
could be even better capable to throw new light on some of the
paradoxes connected with the non-Hermiticity. We are going to pay
attention to one of the most elementary representatives of this
class which is just a simplified form of the potential which we
already analyzed in our older, purely numerical study \cite{DW}
(cf. also \cite{Weigert}). As long as all the similar models
combine solvability (i.e., simplicity) with flexibility
(controlled by several real parameters) we believe that there are
good chances of an explicit confirmation of their full
compatibility with the standard quantum theory. Here we are going
to perform the first steps in this direction.

\section{A Zoo of non-Hermitian Hamiltonians}

Our present study is motivated by the fact that in the light of
our introductory comment the use of non-Hermitian Hamiltonians
sounds like a paradox. Still, in the literature one may notice an
emergence of various sophisticated possibilities of making many of
these models compatible with the concepts of the physical
observability \cite{Geyer,Wang}, with the standard principle of
correspondence \cite{Exper,Kretschmer} or with an introduction of
relativistic kinematics \cite{july} etc.

\subsection{Quasi-Hermitian Hamiltonians}

In a preliminary remark let us remind the reader that even many
finite-dimensional non-Hermitian matrices $A\neq A^\dagger$ may be
re-interpreted as Hermitian after one re-defines the scalar
product of the two complex $N-$dimensional vectors $\vec{f}$ and
$\vec{g}$ accordingly,
 \ben
 \left [
  \left ( \vec{f},\vec{g} \right )_{(I)}= \sum_{n=1}^N\,f^*_n\,g_n
  \equiv \langle f|g\rangle
 \right ]\ \
 \longrightarrow \ \
 \left [
 \left ( \vec{f},\vec{g} \right )_{(\mu)}= \sum_{n=1}^N
 \left (
  \sum_{m=1}^N
 \,f^*_m\,\mu_{m n}\,\right )\,g_n
 \equiv
 \langle \langle f|g\rangle
 \right ]\,.
 \label{jednmi}
 \een
In the other words, one may simply abandon the most common
definition of the bra vector $\langle f|$ (using the mere
operations of the transposition and complex conjugation) and
replace it by its more general alternative with the symbol
$\langle \langle f|$ dependent on the pre-selected metric
$\mu=\mu^\dagger>0$.

The same generalization will also work for Hilbert spaces with $N
= \infty$ after a few necessary but entirely straightforward
additional technical qualifications \cite{Geyer,Kretschmer}. One
only has to keep in mind that any modification of the metric $I
\to \mu$ specifies a different, $\mu-$dependent Hermitian
conjugation mapping $A \to A^\ddagger$,
 \be
 A^\ddagger = \left (
 \mu\,A\,\mu^{-1}
 \right )^\dagger\,,
 \ \ \ \ \ \ \ \ \ \ \mu=\mu^\dagger\,.
 \label{konjugace}
 \ee
This means that the class of the corresponding ''Hermitian"
matrices $A= A^\ddagger$ varies with $\mu$. In order to avoid
confusion, let us call the matrices $A= A^\ddagger$ {\em
quasi-Hermitian} (= the name introduced in ref. \cite{Geyer})
whenever the positive definite metrics $\mu$ becomes different
from the trivial identity operator $I$.

Any such a specific metric operator $\mu>0$ will be denoted by the
symbol $\eta$ in what follows. We must be careful with this
convention since Dirac \cite{Dirac} used the same symbol $\eta$
without requiring its positivity. This means that he spoke about a
pseudo-metric $\mu$, very well exemplified by the indefinite Pauli
matrix occurring in the paper by Feshbach and Villars \cite{FV}
where $\mu=\sigma_3 \bigotimes I$. For this reason, authors of
ref. \cite{Geyer} introduced another symbol for the metric ($\eta
\to T$). Their choice proved unfortunate as it causes a highly
undesirable confusion with the time reversal symbol ${\cal T}$ so
that recently, Ali Mostafazadeh \cite{sqwmetric,alione}
recommended the return to $\eta$ with a specific subscript, $\eta
\to \eta_+$. A less clumsy notation was preferred in refs.
\cite{pseudounitary} or \cite{BBJ} where one starts from certain
auxiliary parity-type indefinite operator ${\cal P}$ and
constructs then the necessary  alternative, ``physical" metrics as
a positive-definite product $\eta \equiv {\cal QP}$ (where ${\cal
Q}$ denotes the so called quasiparity \cite{ptho}) or $\eta \equiv
{\cal CP}$ (with the operator of charge ${\cal C}$ \cite{Wang}),
respectively.

\subsection{Non-quasi-Hermitian time-evolution generators}

Beyond the ``mathematically safe" domain of quantum mechanics as
covered by the majority of the current textbooks one may find a
fairly broad ``grey zone" of models which do not seem to comply
with all the postulates of the theory. One of the illustrations
may be found in relativistic quantum mechanics \cite{Greiner}
where several consequences of the time-evolution laws seem to be
in conflict with the standard (e.g., probabilistic) postulates.
For a brief explicit illustration of the possible difficulties
encountered in similar cases, let us just recollect the exactly
solvable Klein-Gordon model of ref. \cite{KGsusy},
 \ben
 \
 \left( i\,\p_t\right )^2 \Psi^{(KG)}(x,t)
 = \hat{H}^{(KG)}\,\Psi^{(KG)}(x,t)\,,
 \label{jednar}\ \ \ \ \ \  \ \
 H^{(KG)} = -\p^2_x + m^2(x)
 \een
where
 \be
 m^2(x)=m^2_0+\frac{B^2-A^2-A\omega}{\cosh^2\omega x}
 +\frac{B(2A+\omega) \sinh\,\omega x}{\cosh^2\omega x}\,.
 \label{solvable11}
 \ee
Due to its exact solvability the model possesses the bound-state
energies in closed form,
 \ben
 E_n^{(\pm)} = \pm \sqrt{m^2_0-(A-n\omega)^2}
 , \ \ \ \ \ \ n = 0, 1, \ldots, n_{max},
 \ \ \ \ n_{max}=entier[A/\omega]\,.
 \label{spektrsc}
 \een
In the strong-coupling regime with $0 < m_0 < A$, a few low-lying
states suffer a collapse and acquire, formally, complex energies,
i.e., $Im(E_0^{(\pm)}) \neq 0$, $Im(E_1^{(\pm)}) \neq 0$ etc. We
have certainly left the domain of quantum mechanics.

Once we stay within the interval of $0 < A < m_0$, the strength of
the force remains weak and we have the well-behaved eigenvalues
and bound states at all the admissible indices $n$. The
time-evolution of the system is generated by the ``non-Hermitian
Hamiltonian" of ref. \cite{KGsusy}. We may abbreviate
$\varphi_1^{(FV)}(x,t) = i\,\p_t\,\Psi^{(KG)}(x,t)$ and
$\varphi_2^{(FV)}(x,t)=\Psi^{(KG)}(x,t)$ and arrive at the
standard Schr\"{o}dinger-like time-evolution equation
 \ben
 i\,
 \p_t
 \left (
 \ba
 \varphi_1^{(FV)}(x,t)\\ \varphi_2^{(FV)}(x,t)
 \ea
 \right )
=
 \hat{h}^{(FV)}\,
  \left (
 \ba
 \varphi_1^{(FV)}(x,t)\\ \varphi_2^{(FV)}(x,t)
 \ea
 \right ), \ \ \ \ \ \ \ \
 \hat{h}^{(FV)}=\left (
 \begin{array}{cc}
 0&\hat{H}^{(KG)}\\
 1&0
 \ea
 \right )
 \,.
 \label{jdvar}
 \een
where the relativistic Feshbach-Villars-type Hamiltonian
$\hat{h}^{(FV)}$ is not only manifestly non-Hermitian,
 \ben
 \left [\hat{h}^{(FV)} \right ]^\dagger
  = {\cal P}\,\hat{h}^{(FV)}\,{\cal P}^{-1}, \ \ \ \ \ \ \ \
 {\cal P}=\left (
 \begin{array}{cc}
 0&{I}\\
 {I}&0
 \ea
 \right )
 \,,
\label{neheri}
 \een
but also non-quasi-Hermitian since $ \det {\cal P} = -1$. As a
``candidate for the metric" the operator ${\cal P}$ is
unacceptable because it is not positively definite.

In a way coined by Ali Mostafazadeh \cite{Ali}, all the operators
$A=A^\ddagger$ with an indefinite $\mu = {\cal P}$ may be called
{\em ${\cal P}-$pseudo-Hermitian}, assuming only that the
``pseudo-metric" ${\cal P}$ in our Hilbert-Krein space
\cite{Langer} remains invertible. Keeping in mind the paramount
importance of the Klein-Gordon equation we probably should accept
and tolerate the occurrence of the similar time-evolution
generators in physics. Recently, their use became quite useful and
popular in statistical physics \cite{Ahmed} and also beyond the
domain of quantum theory~\cite{Uwe}.

\subsection{${\cal PT}-$symmetric Hamiltonians}

Our choice of the symbol ${\cal P}$ was equally strongly inspired
by its role of a pseudo-metric {\em and} by its coincidence with
the specific parity operator in the Bender's and Boettcher's
pioneering letter \cite{BB}. They paid attention to the bound
states in complex potentials on a line,
 \be
 V(x)\sim (i\,x)^{2+\delta}, \ \ \ \ \ \ \ \ x \in I\!\!R
 , \ \ \ \ \ \ \ \ \delta \in (-1,2)
 \label{unsolvable22}
 \ee
which exhibit an unusual, antilinear parity \& time-reversal
symmetry,
 \be
 {\cal PT} \,V(x)= V(x)\,{\cal PT}
 \label{PTS}
 \ee
where ${\cal P}$ is parity and the symbol ${\cal T}$ denotes the
time-reflection operator \cite{erratum}. Under the condition
(\ref{PTS}) the real bound-state energies may be shown to emerge
for an unexpectedly broad range of parameters \cite{DDT}.

Although the symmetry (\ref{PTS}) may be found in older literature
(cf. refs. \cite{BG} and \cite{Alvarez} working with $\delta=2$
and $\delta=1$, respectively), the importance of the letter
\cite{BB} lies in the courage with which its authors opened the
discussion of the possible sensibility of the study of the
non-quasi-Hermitian models in quantum mechanics. They had to meet
a number of objections a particularly concise sample of which has
been later formulated by R. F. Streater on his web page
\cite{Streater}. Fortunately, a key breakthrough occurred in
several subsequent parallel studies of the ${\cal PT}-$symmetric
Hamiltonians, i.e., of the time-generators $H$ such that $ H
\,{\cal PT} = {\cal PT}\,H$ or, equivalently, $H^\dagger = {\cal
P}\,H\,{\cal P}$ whenever it is assumed that ${\cal P}^{-1}={\cal
P}$. In these studies people paid attention to the exactly
solvable differential-equation models with forces
(\ref{solvable11}) \cite{Quesne} or (\ref{unsolvable22})
\cite{Exper} or to their linear-algebraic diagonal-matrix
reformulations \cite{Most}. It has been clarified that the
``input" pseudo-Hermiticity of a given $H$ may and must be
{complemented} by another, {positively definite} (though, of
course, $H-$dependent) metric $\eta$. Then, the pseudo-Hermitian
Hamiltonians {re-acquire} the quasi-Hermiticity property
 \be
 H^\dagger = \eta\,H\,\eta^{-1}\,, \ \ \ \ \ \ \ \eta=\eta^\dagger>0.
 \label{quasihermiticity}
 \ee
All the requirements are being reduced to an {\em explicit}
assignment of the {\em second}, ``physical" metric $\eta$ to a
given Hamiltonian $H$. One could speak about an ``exotic", ${\cal
PT}-$symmetric version of the standard quantum mechanics.

\section{Point-interaction toy model}


The vast majority of all the applications of ${\cal PT}-$symmetric
quantum mechanics (PTSQM) will rely upon the assumption that for a
given ${\cal P}-$pseudo-Hermitian quantum Hamiltonian $H$ we find
a physical metric operator $\eta$ which is not too complicated.
Most often, the consistent theory is expected to be based on the
factorized form of the metric $\eta = {\cal CP}$ defined in terms
of an {involutive} symmetry ${\cal C}= {\cal C}^{-1}$ (= ``charge"
\cite{BBJ,Wang,Jones} or ``quasi-parity" \cite{ptho,Quesne,Geza}).

The sufficiently simple form of ${\cal C}$ or $\eta$ may only be
assigned to a sufficiently simple ``input" operator $H$. In this
sense, all the ${\cal PT}-$symmetric models using point
interactions \cite{Kurasov,DW,Weigert} represent one of the most
natural playgrounds for the explicit constructions of $\eta$ or
${\cal C}$. In such a context we reported \cite{DW} a few purely
numerical results of the study of a one-dimensional
Schr\"{o}dinger equation
\begin{equation}
\left[ -\,\frac{d^{2}}{dx^{2}} + V(x) \right ] \,\psi (x)=E\psi
(x) \,\ \ \ \ \ \  \psi(\pm L)=0
 \label{schrod}
\end{equation}
with a double-well point-interaction ${\cal PT}-$symmetric
potential $V(x)$ with complex couplings. In our present,
non-numerically oriented continuation of this study let us first
perform a trivial re-scaling of the interval of coordinates which
replaces the usual large though fixed and finite cut-off parameter
$L \gg 1$ by the more comfortable value $L= 1$. Secondly, let us
slightly simplify the potential of ref. \cite{DW} to a mere
two-parametric ${\cal PT}-$symmetric double-well model
 \be
 V(x)=
 -i\,\xi \,\delta(x+a) +
  i\,\xi \,\delta(x-a)\,
 \ \ \ \ \ \ \ x \in (-1,1)
 \label{potik}
 \ee
where the potential is reduced to the mere pair of the delta
functions proportional to a purely imaginary coupling of the size
$\xi$ and located at a distance measured by the variable $ a \in
(0,1)$.

\subsection{Construction of the exact solutions}


We have to solve eq. (\ref{schrod}) with the Dirichlet boundary
conditions $\psi(\pm 1)=0$ and with the piecewise-constant
potential (\ref{potik}) equivalent to the two matching constraints
 \be
 \frac{d}{dx} \psi(\pm a+0)-
 \frac{d}{dx} \psi(\pm a-0)=\pm i\,\xi\,
 \psi(\pm a)\,.
 \label{matching}
 \ee
This task will be facilitated by the observation [based on the
inspection of eq. (\ref{matching}) at small $\xi$] that our
potential (\ref{potik}) represents a smooth perturbation added to
the solvable (and safely Hermitian) bound-state problem in an
infinite square well which possesses the well known and safely
real discrete and positive spectrum. This means that the proof of
the reality and positivity of the energies $E_n$ remains trivial
in the weak-coupling regime. We may assume that $\xi \in (0,
\xi_{crit})$, being aware that the quantity $\xi_{crit}$ may vary
with both the shift $a$ and excitation $n$. This makes the
explicit estimates of the maximal allowed coupling $\xi_{crit}$
important. Their determination will be discussed later. Now, we
only put $E=\kappa^2$ and recollect the usual PTSQM normalization
convention
 \ben
 \psi(x) = \psi_S(x) + i\,\psi_A(x)\,, \ \ \ \ \psi_S(x) =
 \psi_S^\ast(x)=\psi_S(-x)
 \,, \ \ \ \ \psi_A(x) = \psi_A^\ast(x)=-\psi_A(-x)\,
 \een
which significantly facilitates an ansatz for the wave functions,
 \be
 \psi(x)= \left \{
 \begin{array}{ll}
 \psi_L(x) =
 (\alpha - i\,\beta)\,\sin \kappa(x+1), \ \ & x \in (-1,-a)\\
 \psi_C(x) =
 \gamma\,\cos \kappa x + i\,\delta \,\sin \kappa x, \ \ & x \in (-a,a)\\
 \psi_R(x) =
 (\alpha + i\,\beta)\,\sin \kappa(-x+1), \ \ & x \in (a,1).
 \ea
 \right .
 \label{ansatz}
 \ee
Together with the formula for their derivatives,
 \be
 \psi'(x)= \left \{
 \begin{array}{ll}
 \psi'_L(x) =\kappa\,
 (\alpha - i\,\beta)\,\cos \kappa(x+1), \ \ & x \in (-1,-a)\\
 \psi'_C(x) =-\kappa\,
 \gamma\,\sin \kappa x + i\,\kappa\,\delta \,\cos \kappa x,
 \ \ & x \in (-a,a)\\
 \psi'_R(x) =-\kappa\,
 (\alpha + i\,\beta)\,\cos \kappa(-x+1), \ \ & x \in (a,1)
 \ea
 \right .
 \ee
both these formulae enter the matching constraint (\ref{matching})
and transforms it into the following four-by-four matrix equation,
 \be
 \left (
 \begin{array}{cccc}
 \sin \kappa (1-a)& 0& -\cos \kappa a& 0\\
 0& \sin \kappa (1-a)& 0& -\sin \kappa a\\
  -\cos \kappa (1-a)& \xi\,\kappa^{-1}\,\sin \kappa (1-a)
   & \sin \kappa a& 0\\
 \xi\,\kappa^{-1}\,\sin \kappa (1-a)
 & \cos \kappa (1-a)& 0& \cos \kappa a
  \ea
 \right )
 \left ( \ba
 \alpha\\
 \beta\\
 \gamma\\
 \delta
 \ea
 \right ) = 0\,.
 \label{match}
 \ee
It possesses nontrivial solutions if and only if its secular
determinant ${\cal D}(\kappa)$ vanishes,
 \be
 {\cal D}(\kappa)
 \equiv
 -\frac{1}{2}
 \left \{
 \sin\, 2\kappa +
 \frac{\xi^2}{\kappa^2}\,
 \sin \,2\kappa a\,\cdot
 \sin^2 [\kappa(1- a)]
 \right \}
  =0
 \,.
 \label{side}
 \ee
The sequence of the nodal zeros $\kappa_n=\kappa_n(a,\xi)> 0$ of
this trigonometric function determines the spectrum of the
energies $E_n=\kappa^2_n>0$. The coefficients $\alpha, \beta,
\gamma$ and $\delta$ define the wave function and are also given
by the nice and compact formulae, say,
 \be
 \gamma= \frac{\xi\,\sin \kappa a\,\sin \kappa (1-a)}
 {\kappa\,\cos \kappa}\,\delta\,,
 \ \ \ \ \
 \beta=\frac{\sin \kappa a}{\sin \kappa (1-a)}\,\delta
 \label{osel}
 \ee
 \be
 \delta= -\frac{\xi\,\cos \kappa a\,\sin \kappa (1-a)}
 {\kappa\,\sin \kappa}\,\gamma\,,
 \ \ \ \ \
 \alpha=\frac{\cos \kappa a}{\sin \kappa (1-a)}\,\gamma
 \label{stin}
 \ee
with an appropriate choice of a convenient normalization.

\subsection{Onset $\xi_{crit}$ of the spontaneous
 ${\cal PT}-$symmetry breakdown}

The compact solvability of our matching conditions enables us to
determine the range of  couplings $\xi$ for which all the roots
$\kappa_n=\kappa_n(a,\xi)$ of our secular eq. (\ref{side}) remain
real.  We shall see that it happens within fairly large intervals
of $\xi \in [0,\xi_{crit}(a)]$, at the upper end of which one
encounters the generic merger and subsequent complexification of
some of the low-lying energy pairs \cite{Langer}.

\subsubsection{The reality of the energies in the weak-coupling
regime}

In the PTSQM context it is usually emphasized that after the
transition through the point $\xi_{crit}(a)$ (which is usually
called an ``exceptional point" \cite{Heiss}), the complexification
of the energies is accompanied by the (so called spontaneous)
breakdown of the ${\cal PT}-$symmetry of the wave functions. The
rigorous study of this phenomenon proves significantly facilitated
in our specific, solvable point-interaction model. Firstly we may
drop the obvious zero root $\kappa=0$ of eq. (\ref{side}) as
spurious, giving just the trivially vanishing wave functions.
Secondly, we may start our considerations from the domain of the
very small $\xi \ll 1$ where it is easy to observe that the well
known square-well roots are only slightly perturbed during the
growth of $\xi$,
 \be
 \kappa_n(a,\xi) \approx \frac{n}{2}\,\pi +
 {\cal O}(\xi^2), \ \ \ \ \ n = 1, 2,
 \ldots\,.
 \label{roots}
 \ee
With the further growth of $\xi$ we notice that the second term in
eq. (\ref{side}) will be most influential at the smallest energy
levels $\kappa_n^2$. The main uncertainty in an estimate of its
effect lies in the $a-$dependence of its trigonometric factor.

\subsubsection{An explicit formula for $\xi_{crit}(1/2)$}

The mechanism of the spontaneous breakdown of the ${\cal
PT}-$symmetry will be best visible in the most symmetric case with
the half-unit displacement $a=1/2$ of our $\delta-$functional
interaction points in eq. (\ref{potik}). In the secular equation
 \be
 \sin\, \kappa \,
 \left [ \cos\,\kappa+
 \frac{\xi^2}{4\,\kappa^2}\,
 (1-\cos\,\kappa)
 \right ]=0, \ \ \ \ \ \ \ a=\frac{1}{2}
 \label{pul}
 \ee
the odd-parity square-well roots [= those with even subscripts $n$
in the zero-order eq. (\ref{roots})] will not move with the growth
of $\xi$ at all, due to the $\xi-$independence of the first factor
in eq. (\ref{pul}). In contrast, the $\xi-$dependence of the
odd$-n$ series of the roots will be nontrivial and controlled by
their implicit definition (\ref{pul}) or, after a simplification,
by the equation
 \be
 \cos\,\kappa = -\frac{\mu^2}{\kappa^2-\mu^2}, \ \ \ \ \ \
 \mu = \frac{\xi}{2}\,.
 \label{impro}
 \ee
Such a picture is extremely transparent because the
right-hand-side curve is a smooth and growing function of $\kappa
\geq 0$ with a single pole at $\kappa=\mu$. Its values are larger
than one at the smallest $\kappa \leq \mu$ where it cannot produce
any roots since we must have $|\cos \,\kappa| \leq 1$. At all the
larger $\kappa > \mu=\xi/2$ the latter curve is negative. Of
course, it should not lie below the minus-one lower limit so that
we may set an improved estimate of the roots as $\kappa>
\sqrt{2}\,\mu=\xi/\sqrt{2}$. This is a constraint imposed, in
effect, not only upon the energies $\kappa^2$ in a given potential
but also upon the allowed measure $\xi$ of the manifest
non-Hermiticity at a given energy.

An inspection of eq. (\ref{impro}) reveals that the growth of
$\xi$ implies that the first two $\xi-$dependent energy levels
$E_1=\kappa_1^2(1/2,\xi)$ and $E_3=\kappa_3^2(1/2,\xi)$  move
towards each other until they coincide at a certain critical
strength $\xi_{crit}=2\mu_0$. We have $\kappa_1(1/2,\xi_{crit})
=\kappa_3(1/2,\xi_{crit})=\nu_0>\pi$ at this point. In the other
words, the left- and right-hand-side curves in eq. (\ref{impro})
will touch at a point which is determined by the two coupled
equations
 \be
 \cos\,\nu_0 = -\frac{\mu^2_0}{\nu^2_0-\mu^2_0}, \ \ \ \ \ \
 -\sin\,\nu_0 = \frac{2\,\mu^2_0\,\nu_0}{\left (
 \nu^2_0-\mu^2_0 \right )^2}\ .
 \label{coupled}
 \ee
We eliminate
 \ben
 \mu_0^2=-2\,\nu_0\,\frac{\cos^2\,\nu_0}{\sin\,\nu_0}
 \een
and get the reduced equation for the remaining unknown,
 \ben
 \nu_0=2\,\tan\,\frac{1}{2}\,\nu_0\,\cos\,\nu_0.
 \een
We may find its root $\nu_0\approx 3.874366817$ numerically, at
$\mu_0\approx 2.529882472$, i.e., at the critical strength
$\xi_{crit}\approx 5.059764944$.

The touching-point $\xi_{crit}$ will grow with the excitation. We
may conclude that the spectrum remains real at {\em all} the
couplings $\xi \leq \xi_{crit}\approx 5.06$. This means that the
${\cal PT}-$symmetry  becomes spontaneously broken and some of the
energies complexify {\em only} beyond this critical strength of
the non-Hermiticity.

\subsection{Unavoided crossings of the energy levels}

\subsubsection{Facilitated construction at $a=1/2$}

The touching point of the negative branch of the cosine with an
increasing right-hand-side hyperbolic curve of eq. (\ref{impro})
is displaced to the right of the first excited level,
$\nu_0>\kappa_2(1/2, \xi) =\pi$. The latter level was generated by
the sine part of secular equation (\ref{pul}) and coincides with
the position of the minimum of the cosine curve. We may conclude
that as long as we had the ground-state level
$\kappa_1(1/2,0)=\pi/2$ at $\xi = 0$, the continuous and growing
function $\kappa_1(1/2,\xi)$ of $\xi$ crosses the line of
$\kappa_2(1/2,\xi)$ at some particular coupling strength
$\xi_{1\otimes 2}$ which lies somewhere inside the interval
$(0,\xi_{crit})$.

The existence of the similar points of an unavoided crossing of
the real energy levels has been spotted in several other solvable
models \cite{ptho,morse,fragil}. In the present $a=1/2$
illustrative example these unavoided crossings will involve all
the pairs of the real levels $\kappa_{4m-3}(1/2,\xi)$ and
$\kappa_{4m-2}(1/2,\xi)$. Their crossings are unique and will take
place at the special strengths $\xi=\xi_{(4m-3)\otimes (4m-2)}$
defined by eq. (\ref{pul}) which leads to the elementary formula
 \be
 \xi_{(4m-3)\otimes (4m-2)}=\sqrt{2}\,\pi\,(2m-1)\,,
 \ \ \ \ \ \ \ m = 1,
 2, \ldots
 \,.
 \label{pulct}
 \ee
At these crossing points, our non-Hermitian Hamiltonian ceases to
be diagonalizable and remains only block-diagonalizable,
containing the so called Jordan canonical irreducible triangular
submatrices \cite{Azizov}. Whenever we wish to avoid similar
subtleties, it suffices to assume that $\xi < \min \left (
\xi_{(4m-3)\otimes (4m-2)} \right ) = \xi_{1\otimes
2}=\sqrt{2}\,\pi\approx 4.442882938 $.

\subsubsection{Approximate constructions in the
vicinity of $a=1/2$}

The remarkable factorization (\ref{pul}) of our general secular
equation hints that in a small vicinity of $a=1/2$ one could still
expect simplifications. Introducing a small measure of
perturbation $\sigma$ let us write $a=1/2+\sigma/(2\kappa)$ and
return once more to our exact secular eq. (\ref{side}) which
reads, in the new notation,
 \ben
 \sin\, 2\kappa +
 \frac{\xi^2}{2\,\kappa^2}\,
 \sin \,
 \left (
 \kappa+\sigma
 \right )
 \cdot
 \left [
 1-
 \cos \,
 \left (
 \kappa-\sigma
 \right )
 \right ]
 =0\,.
 \een
In the first two orders of $\sigma$ the second term has the form
of the product
  \ben
 \frac{\xi^2}{2\,\kappa^2}\,
  \left (
  \sin \, \kappa+
  \sigma\,\cos \, \kappa-\frac{1}{2}\,
  \sigma^2\sin \, \kappa
 \right )
 \cdot
 \left (
 1-
 \cos \,
 \kappa
 -
  \sigma\,\sin \, \kappa+\frac{1}{2}\,
  \sigma^2\cos \, \kappa
 \right )
 \een
so that we may study the following approximation of the secular
equation,
  \be
  \sin\, 2\kappa+
 \frac{\xi^2}{2\,\kappa^2}\,
  \left [
  \sin \, \kappa\,
  \left (
 1-
 \cos \,
 \kappa
 \right )
 -\sigma\, \left (
 1-
 \cos \,
 \kappa
 \right )
 -\frac{1}{2}\,\sigma^2
 \sin \, \kappa
 \right ]
   =
  {\cal O} \left (\sigma^3 \right )\,.
  \label{approxi}
 \ee
First thing we notice is that at $\sigma \neq 0$ the
factorizability of eq.~(\ref{pul}) is lost so that one cannot
factor the term $\sin\kappa$ out. This means that all the
sine-generated roots $\kappa_2(a,\xi),\, \kappa_4(a,\xi),\,
\kappa_6(a,\xi),\, \ldots $ (which were $\xi-$independent at
$\sigma=0$) may become $\xi-$dependent even near the square-well
regime with very small $\xi \approx 0$.

Within our precision of $1+{\cal O} \left (\sigma^3 \right )$ let
us employ the partially factorized approximate secular equation
(\ref{approxi}),
  \ben
  \sin \frac{1}{2} \kappa\,
  \left \{
  4\,\cos \frac{1}{2}
   \kappa\,\cos\, \kappa+
 \frac{\xi^2}{2\,\kappa^2}\,
  \left [
  2\,\sin \, \kappa\,
  \sin \frac{1}{2} \kappa
 -2\,\sigma\,\sin \frac{1}{2} \kappa
 -\sigma^2\,\cos \frac{1}{2}
   \kappa
 \right ]
 \right \}
   =0\,.
 \een
In its light the subset of the roots $\kappa_4(a,\xi),\,
\kappa_8(a,\xi),\, \kappa_{12}(a,\xi),\, \ldots $ still remains
more or less $\xi-$independent at the sufficiently small $\sigma
\sim a-1/2$. For the complementary family with the safely
non-vanishing $ \sin \frac{1}{2} \kappa$ we get, in the first
nontrivial approximation, the implicit definition
  \ben
    \cot \frac{1}{2}
   \kappa\,\cos\, \kappa+
 \frac{\xi^2}{4\,\kappa^2}\,
  \left [
  \sin \, \kappa
 -\sigma
  \right ]
   =0\,
 \een
which indicates that all the roots $\kappa_2(a,\xi),\,
\kappa_6(a,\xi),\, \kappa_{10}(a,\xi),\, \ldots$ which annihilated
$\cos ({\kappa }/{2})$ at $\alpha=1/2$ (i.e., at $\sigma=0$) will
now grow with $\sigma$. As one can easily demonstrate, they will
all move in proportion to the factor ${\xi^2}/4\kappa^2$ at small
$\xi$.

We may conclude that although the study of our toy model in the
domain where $a -1/2$ does not vanish is perceivably more
complicated in the technical sense, a straightforward perturbative
approach to a systematic derivation of the ${\cal O}(\sigma^k)$
corrections is still feasible. Hence, we may recommend it as a
complement or alternative to the less sophisticated numerical
constructions.

Skipping the details we shall now return to $a=1/2$ and show how
our model may be interpreted as a mere small perturbation of its
Hermitian square-well $\xi \to 0$ limit.

\section{The excitation-dependence of the solutions}

Our construction of the solutions need not necessary be only based
on the purely numerical determination of the roots $\kappa_n$ from
transcendental equations. Alternatively, perturbative methods are
also easy to apply, especially when $\kappa_n={\cal O}(n)$ is
large.

\subsection{A power-series analytic representation
of the energies}

Whenever the value of $\xi$ stays safely below the ${\cal
PT}-$symmetry breakdown boundary $\xi_{crit} \approx 5$, all the
roots $\kappa_n(a,\xi)$ of our secular equation (\ref{side}) with
$n=1,2,\ldots$  converge to their compact square-well limits
$\kappa_n(a,0)=n\pi/2$ not only when we move $\xi \to 0$ but also
for the growing $n \to \infty$. In order to describe this
phenomenon in more detail, let us return back to the secular
equation (\ref{pul}) where we may skip the trivial case (=
even-subscripted closed solutions $\kappa_{2m}(1/2,\xi)\equiv
m\pi$ with $m=1,2,\ldots$) and introduce the following new
variables for the odd-subscripted roots,
 \be
 \kappa_{4m-3}=\frac{(4m-3)\,\pi}{2}+x_m, \ \ \ \
 \kappa_{4m-1}=\frac{(4m-1)\,\pi}{2}-y_m, \ \ \ \ m = 1, 2,
 \ldots\,.
 \label{medium}
 \ee
This simplifies the secular sub-equation (\ref{impro}) which forms
the series of equations
 \be
 \sin x = \frac{1}{L^2(m,x)-1},
 \ \ \ \ \ \ \ L(m,x)=\frac{(4m-3)\,\pi+2x}{2\,\mu},
  \ \ \ \ m = 1, 2, \ldots,
 \label{improva}
 \ee
for the unknown $x=x_m$ and another, slightly different series for
$y=y_m$,
 \be
 \sin y = \frac{1}{U^2(m,y)-1},
 \ \ \ \ \ \ \ U(m,y)=\frac{(4m-1)\,\pi-2y}{2\,\mu},
  \ \ \ \ m = 1, 2, \ldots\,.
 \label{improbe}
 \ee
The roots remain bounded, $x_n< \pi$ and $y_m<\pi$, but they
depend on both the coupling $\mu=\xi/2$ and index $m$. As long as
equations (\ref{improva}) and (\ref{improbe}) are not too
dissimilar, we may treat them as two special cases of a single
equation,
 \be
 \sin z =
 \left [
 \left (
 \frac{1}{\lambda\mu}-\frac{z}{\mu}
 \right )^2
 -1
 \right ]^{-1},
 \ \ \ \ \ \ \
 z \ =\  \left \{
 \ba
 x_m\\
 y_m
 \ea
 \right .\,,
 \ \ \ \ \ \ \
 \lambda \ =\
 \left \{
 \ba
 -\frac{2}{(4m-3)\pi}\\
 +\frac{2}{(4m-1)\pi}
 \ea
 \right .\,.
 \label{probant}
 \ee
Once $\lambda$ is small, it is trivial to solve it in closed form,
say, by iterations,
 \be
 z={\lambda}^{2}{\mu^{2}}+{\lambda}^{4}{\mu^{4}}+{
 {2}}\,{\lambda}^{5}{\mu^{4}}+\frac{7}{6}\,{\lambda}^{6}{\mu^{6}}+
 {6}\,{\lambda}^{7}{\mu^{6}}+\left (7+
 \frac{3\,{\mu^{2}}}{2}\right )
 {\lambda}^{8}{\mu^{6}}+{\cal R}(\lambda,\mu)\,.
 \label{enepert}
  \ee
Here, the $ {\cal O} \left ( {\lambda}^{9}{\mu^{8}} \right )$
remainder has transparent structure
 \ben
 {\cal R}(\lambda,\mu)=
 {\frac {40}{3}}\,\lambda^{9}{\mu^{8}}+\left ({36}+{\frac
 {83{\mu^{2}}}{40}}\right ){\lambda }^{10}{\mu^{8}}+
 \left (30+
 \frac {80\,{\mu}^{2}}{{{3}}}
  \right )
 {\lambda}^{11}{\mu^{8}}
 +{\cal O} \left (
 {\lambda}^{13}{\mu^{10}}
 \right )
 \een
and admits the further higher-order improvements if needed. Note
that this series is not changing signs which means that the
function itself is monotonic and growing with $\lambda > 0$. It
behaves like a perturbation series in both $\lambda$ (inversely
proportional to the excitation number $m$) and $\mu=\xi/2$ (the
coupling) {\em simultaneously}.

\subsubsection{A check of consistency}

We may compare the rate of the growth of $x_m$ and $y_m$ with an
increase of the coupling $\mu$. This requires a removal of the
difference in the $m-$dependence of the respective $\lambda$s, via
an obvious transition to the same, shared $m-$dependent factor
$\tau$,
 \ben
 z \ =\  \left \{
 \ba
 x_m\\
 y_m
 \ea
 \right .
 \ \ \ \Longrightarrow \ \ \
 \pi\,\lambda \ =\
 \left \{
 \ba
 -\frac{2}{(4m-3)}
 = -\frac{\tau}{1-\tau/2}
 \\
 +\frac{2}{(4m-1)}= +\frac{\tau}{1+\tau/2}
 \ea
 \right .\,,
 \ \ \ \ \ \ \tau =\tau(m)=\frac{1}{(2m-1)}>0
 \,.
 \een
The use of the new variable defines the functions $x_m=x(\tau)$
and $y_m=y(\tau)$ in such a way that $x(\tau)=y(-\tau)$, i.e.,
{\em both} of them are given by the {\em same} series in $\tau$,
 \ben
 z=
\left \{
 \ba
 x_m\\
 y_m
 \ea
 \right \}\ =
{\frac {{\mu}^{2}}{{\pi}^{2}}}{\tau}^{2}\pm {\frac
{{\mu}^{2}}{{\pi}^{2}}}{ \tau}^{3}+\left ({\frac
{{3\mu}^{2}}{{4\pi}^{2}}}+{\frac {{\mu}^{4}}{{ \pi}^{4}}}\right
){\tau}^{4} \pm \left ({\frac {{\mu}^{2}}{{2\pi}^{2}}}-2 \,{\frac
{{\mu}^{4}}{{\pi}^{5}}}+2\,{\frac {{\mu}^{4}}{{\pi}^{4}}} \right
){\tau}^{5}+\ldots\,.
 \een
As expected, $x(\tau)$ grows more rapidly with $\tau$ so that we
arrived at an independent, non-trigonometric proof that $x_m> y_m$
for all $m$ for which these series converge. Unfortunately, in
contrast to our previous formula (\ref {enepert}) which offers a
systematic improvement of convergence  with respect to {\em both}
the increasing couplings $\mu=\xi/2$ {\em and} the decreasing
excitations $m \ll \infty$, the latter two power series in $\tau$
have an error term ${\cal O}(\mu^2)$ so that they remain less
useful unless the coupling $\mu$ remains very small.

\subsection{Matrix elements and wave functions}

As long as we work just with the four-dimensional matrix problem
(\ref{match}), it makes sense to derive the perturbed wave
functions directly from its $a=1/2$ version with abbreviations
${S} = \sin (\kappa/2)$ and ${C} = \cos (\kappa/2)$,
 \be
 \left (
 \begin{array}{cccc}
 {S}& 0& -{C}& 0\\
 0& {S}& 0& -{S}\\
  -{C}& \xi\,\kappa^{-1}\,{S}
   & {S}& 0\\
 \xi\,\kappa^{-1}\,{S}
 & {C}& 0& {C}
  \ea
 \right )
 \left ( \ba
 \alpha\\
 \beta\\
 \gamma\\
 \delta
 \ea
 \right ) = 0\,.
 \label{matchversionb}
 \ee
An analysis of such a compactified Schr\"{o}dinger equation must
be performed separately in the two or three different domains of
$\kappa$.

\subsubsection{Special cases of the levels when $S=0$ or $C=0$}

Once we admit that $S=0$, we infer that $\alpha=\gamma=0$ and
$\beta=-\delta \neq 0$ while $C = \pm 1$. Wave functions $ \psi(x)
\sim \sin \kappa x$ vanish precisely at the points of interaction
and so they ``do not see it" and remain independent of $\xi$. The
whole family of these $ \psi(x)$ coincides, at all the real
non-Hermiticites $\xi$, with the pure square-well states
$\psi_4(x), \psi_8(x), \ldots$ with subscripts $n \equiv 0 \,({\rm
mod}\ 4)$.

In another special case of the matrix elements such that $C=0$,
wave functions may be re-normalized to having antisymmetric real
part,
 \be
 \psi(x)= \left \{
 \begin{array}{ll}
 \psi_L(x) =
  \sin \kappa(x+1), \ \ & x \in (-1,-1/2)\\
 \psi_C(x) =
 -i\xi\kappa^{-1}
 \,\cos \kappa x -\sin \kappa x, \ \ & x \in (-1/2,1/2)\\
 \psi_R(x) =
  \sin \kappa(x-1), \ \ & x \in (1/2,1).
 \ea
 \right .
 \ee
At all $\xi$ the corresponding energies remain also equal to the
Hermitian square-well values $E_2, E_6, \ldots$. We may notice
that there exists also a non-trivial symmetric imaginary part of
$\psi(x)$ which vanishes for $|x|>1/2$ and which grows in
proportion to the coupling $\xi$.

\subsubsection{Generic Schr\"{o}dinger equation}

At the generic matrix elements $S\neq 0 \neq C$ our
four-dimensional Schr\"{o}dinger equation~(\ref{matchversionb})
implies the simplification $\delta=\beta$. This may be inserted in
our ansatz (\ref{ansatz}) for wave functions and enables us to
drop the second line in eq. (\ref{matchversionb}) as redundant,
 \be
 \left (
 \begin{array}{ccc}
 {S}& 0& -{C}\\
  -{C}& \xi\,\kappa^{-1}\,{S}
   & {S}\\
 \xi\,\kappa^{-1}\,{S}
 & 2{C}& 0
  \ea
 \right )
 \left ( \ba
 \alpha\\
 \beta\\
 \gamma
 \ea
 \right ) = 0\,.
 \label{matchversionc}
 \ee
This simplifies the derivation of the other two closed formulae,
 \be
 \beta= \delta
 = -\frac{\xi}{2\kappa}\,\tan ({\kappa }/{2})\,\alpha\,,
  \ \ \ \ \ \ \ \ \ \ \
 \gamma=\tan ({\kappa }/{2})\,\alpha\,.
 \label{gza}
 \label{bzea}
 \ee
We are left with the single equation requiring that the secular
determinant vanishes. We may freely choose a normalization
$\alpha=1$ and conclude that the determination of the wave
functions is completed. Their explicit perturbative dependence on
the excitation number $m$ is mediated by formulae (\ref{medium})
and (\ref{enepert}).

\section{The coupling-dependence of the solutions}

Our reduced secular equation (\ref{impro}) contains the
trigonometric functions ${S}$ and ${C}$. In an opposite direction
we may abbreviate $\varrho = \xi/\kappa$ and eliminate  ${S}$ and
${C}$, transforming this secular equation into one of the
following two equivalent new formulae,
 \be
 {S}^2= \frac{1}{2-\varrho^2/2}\,, \ \ \ \ \ \ \
 {C}^2= \frac{1-\varrho^2/2}{2-\varrho^2/2}\,.
 \label{expt}
 \ee
These relations re-define the matrix elements in our
four-dimensional ``perturbed" Schr\"{o}dinger equation
(\ref{matchversionb}) as functions of an alternative new variable
$\varrho$ which remains small in the weak-coupling regime..

\subsection{An illustration with $z=x_m$}

The choice of $z=x_m$ covers the energy levels $E_1$ (= ground
state), $E_5$, $E_9$ etc. They are all slightly greater than their
``unperturbed", square-well $\xi=0$ predecessors.

From our previous analysis of the subset of eigenvalues $E_{4m-3}$
corresponding to $z=x_m$ we infer that below the level crossing,
i.e., for all the not too large couplings $\xi<\xi_{(4m-3)\otimes
(4m-2)}=\sqrt{2}\,\pi\,(2m-1)$ we may take the square roots in eq.
(\ref{expt}) unambiguously. Without any loss of generality (i.e.,
up to an irrelevant overall sign), we shall evaluate our matrix
elements in accordance with the unique recipe
 \be
 {S}= {S}_x(\varrho)=\sqrt{\frac{1}{2-\varrho^2/2}}\ \ >\ \
 {C}= {C}_x(\varrho)=\sqrt{\frac{1-\varrho^2/2}{2-\varrho^2/2}}\ >0\,.
 \label{exptx}
 \ee
In both these functions we may consider the quantity $\varrho$
small and employ their known Taylor expansions,
 \ben
 \sqrt{2}\, {S}_x(\varrho)=1+{\frac {1}{8}}{\varrho}^{2}
 +{\frac {3}{128}}{\varrho}^{4}+{\frac {5}{1024}}
 {\varrho}^{6}+{\frac {35}{32768}}{\varrho}^{8}
 +{\frac{63}{262144}}{\varrho}^{10}+O\left ({\varrho }^{12}\right )
 \een
 \ben
 \sqrt{2}\, {C}_x(\varrho)=1-{\frac {1}{8}}{\varrho}^{2}-{\frac {5}{128}}
 {\varrho}^{4}-{\frac {13}{1024}}{\varrho}^{6}
 -{\frac {141}{32768}}{\varrho}^{8}-{\frac {399}{262144}}
 {\varrho}^{10}+O\left ({\varrho}^{12}\right )\,.
 \een
Another Taylor series carrying a weak-coupling perturbation
expansion character,
 \ben
 {S}_x(\varrho)/{C}_x(\varrho)=1+{\frac {1}{4}}{\varrho}^{2}
 +{\frac {3}{32}}{\varrho}^{4}+{\frac {5}{128}}{\varrho}^{6}
 +{\frac {35}{2048}}{\varrho}^{8}+O\left ({\varrho}^{10}\right )
 \een
enters the formulae (\ref{gza}) which define all the components
$\gamma=\gamma(\varrho)\equiv \delta$ and $\beta=\beta(\varrho)$
of the perturbed wave functions.

\subsubsection{The second special case with $z=y_m$}

The fourth quarter of the set of all eigenvalues comprises the
energies $E_{4m-1}$ which lie below their unperturbed partners and
which correspond to the choice of $z=y_m$. In this case we have to
take the square roots in eq. (\ref{expt}) in accordance with this
modified expectation,
 \be
 {S}= {S}_y(\varrho)=\sqrt{\frac{1}{2-\varrho^2/2}}\ \ >\ \
 -{C}= -{C}_y(\varrho)=\sqrt{\frac{1-\varrho^2/2}{2-\varrho^2/2}}\ >0\,.
 \label{expty}
 \ee
The necessary parallel modification of all the previous $z=x_m$
results and formulae is trivial.

\subsection{Wave functions in dependence on both $\xi$ and $1/m$}

Our wave functions have to be represented in terms of both the
complementary and independent measures of the non-Hermiticity
$\lambda=\lambda(m)$ and $\mu=\xi/2$. The final transition from
the intermediate auxiliary quantity $\varrho$ to this original
pair of expansion parameters profits from the feasibility of the
Taylor-expansion technique. Up to corrections $O\left
({\lambda}^{9}\,{\mu}^{8} \right )$ we have the following final
result for the matrix elements in eq. (\ref{matchversionb}),
 \ben
 \sqrt{2}\, {S}_x[\varrho(\lambda)]\approx
 1+\frac{1}{2}\,{\mu}^{2}{\lambda}^{2}+\frac{3}{8}\,{\mu}^{4}{\lambda}^{4}
 -{\mu}^{4}{\lambda}^{5}+{
\frac{5}{16}}\,{\mu}^{6}{\lambda}^{6}-\frac{5}{2}\,{\mu}^{6}{\lambda}^{7}
+\left({\frac { 35}{128}}\,{\mu}^{8}-\frac{1}{2}\,{\mu}^{6}
\right){\lambda}^{8}\,,
 \een
 \ben
 \sqrt{2}\, {C}_x[\varrho(\lambda)]\approx
1-\frac{1}{2}
\,{\mu}^{2}{\lambda}^{2}
 -\frac{5}{8}\,{\mu}^{4}{\lambda}^{4}+{\mu}^{4}{\lambda}^{5}-{
\frac
{13}{16}}\,{\mu}^{6}{\lambda}^{6}+\frac{7}{2}\,{\mu}^{6}{\lambda}^{7}
+\left(-{ \frac {141}{128}}\,{\mu}^{8}+\frac{1}{2}\,{\mu}^{6}
 \right){\lambda}^{8}\,.
 \een
This converts our Schr\"{o}dinger equation (\ref{matchversionb})
of the form $Q(\lambda,\mu)\,\vec{\psi}=0$ into its power-series
four-by-four matrix version
 \ben
 \left [
 Q(0,0) + \mu^2\lambda^2\,Q_{[1]} + \mu^4\lambda^4\,Q_{[2]} +
  \mu^4\lambda^5\,Q_{[3]} + \ldots
 \right ]\,\vec{\psi}(\lambda,\mu)
 =0\,.
 \een
The standard Rayleigh-Schr\"{o}dinger perturbation theory
\cite{Messiah} might apply and define the sequence of corrections,
order-by-order in both our small parameters. The exact solvability
of our problem facilitates the underlying manipulations and
reduces the abstract algorithm to the explicit formulae
(\ref{bzea}) where the only ingredient we need to know is the
expansion of the single auxiliary factor $S/C$. It is easy to
derive its sixth-order weak-coupling form
 \ben
 \tan \left [\kappa_m\left (\frac{1}{2},2\mu \right )
 \right ]\approx
1+{\mu}^{2}{\lambda}^{2}+\frac{3}{2}\,{\mu}^{4}{\lambda}^{4}
-2\,{\mu}^{4}{\lambda}^{5}+\frac{5}{2}\,{
\mu}^{6}{\lambda}^{6}-8\,{\mu}^{6}{\lambda}^{7}+\left ({\frac
{35}{8}}\,{\mu}^{8}- {\mu}^{6}\right ){\lambda}^{8}\,
 \een
as well as all its further higher-order improvements whenever
needed.

\section{Summary and outlook}

We described an application of the PTSQM formalism to a toy-model
Hamiltonian $H$ with the real energies $E_n=\kappa^2_n$ which
remain non-degenerate for $\xi<\xi_{1\otimes 2}$. The set of the
bound states $|n\rangle$ (\ref{ansatz}) is not orthogonal since
$H^\dagger = {\cal P}\,H\,{\cal P} \neq H$. Still, in our Hilbert
space we may construct a biorthogonal basis using the second set
of the eigenstates of $H^\dagger$ corresponding to the same
spectrum of course. They will be distinguished by the
double-bra/ket symbols (e.g., $\langle\langle n|$ or
$|n\rangle\rangle$, respectively), being most naturally understood
as the ``left eigenstates of $H$" \cite{pseudounitary} defined by
equation
 \ben
 H^\dagger |m\rangle\rangle = E_m\,
 |m\rangle\rangle ={\cal P}\,H\,{\cal P}\,|m\rangle\rangle\,.
 \een
Due to the involutivity of the parity (${\cal P}^2=I$) and due to
the non-degeneracy of the spectrum we know that ${\cal
P}\,|m\rangle\rangle$ must be proportional to $|m\rangle$ so that
we are free to define
 \ben
 |m\rangle\rangle = q_m^*\,{\cal P}\,|m\rangle\,
 \een
using any normalization constant $q_m^*$  (for our future
convenience we write here the star $^*$ which denotes the complex
conjugation). It is also easy to deduce the law of biorthogonality
$\langle\langle n|m\rangle \sim \delta_{mn}$ (hint: subtract the
two suitably pre-multiplied equations $H\,|m\rangle -
E_m\,|m\rangle $ and $\langle\langle n|\,H = E_n\langle\langle
n|$). Once we evaluate all the overlaps $R_n=\langle\langle
n|n\rangle$, we may formally write down the explicit decomposition
of the unit operator,
 \ben
 I =
 \sum_{n=1}^\infty\,|n\rangle\,\frac{1}{R_n}\,\langle\langle  n|=
 \sum_{n=1}^\infty\,|n\rangle\rangle\,\frac{1}{R_n^*}\,\langle
 n|
 \een
as well as the spectral decompositions of our two forms of the
Hamiltonian,
 \be
 H = \sum_{n=1}^\infty\,|n\rangle\,\frac{E_n}{R_n}\,\langle\langle
 n|, \ \ \ \
 H^\dagger = \sum_{n=1}^\infty\,|n\rangle\rangle\,\frac{E_n}{R_n^*}
 \,\langle  n|\,.
 \label{donald}
 \ee
In practice, we determine the values of $R_n$ by calculating
matrix elements of parity which {\em are} all real but which {\em
cannot} be all of the same sign, $\langle n|{\cal P}|n \rangle =
R_n/q_n$.

The key step now lies in making the PTSQM formalism compatible
with the postulates of quantum theory by an introduction of some
new operator $\eta$ of ``the physical" metric in Hilbert space. We
remind the reader that it need not be involutive ($\eta^2 \neq I$
in general \cite{Geyer}) but it must be Hermitian and positive
definite, i.e., $\eta =\eta^\dagger\neq {\cal P}$ (in this sense,
the indeterminate ${\cal P}$ is {\em not} a metric). The main
purpose of the search for such a metric $\eta$ is that it makes
our Hamiltonian quasi-Hermitian and, hence, physical,
 \ben
 H^\dagger = \eta\,H\,\eta^{-1}\,.
 \een
Let us now re-interpret this relation as an equation for $\eta$
which we have to solve. In the light of eq. (\ref{donald}) we have
 \be
  \sum_{n=1}^\infty
 \eta\,\,|n\rangle\,\frac{E_n}{R_n}\,
 \langle\langle n|\ = \
 \sum_{m=1}^\infty\,|m\rangle\rangle\,\frac{E_m}{R_m^*}
 \,\langle  m|\,\eta
 \label{tuhyk}
 \ee
which implies that, in general,
 \ben
 \eta=
 \sum_{m,n=1}^\infty\,|n\rangle\rangle\,M_{nm}\,\langle\langle m|
 \,.
 \een
The backward insertion of such an ansatz in eq. (\ref{tuhyk})
gives, at all $n$ and $m$, the condition $ E_n\,M_{nm} = M_{nm} \,
E_m$ with the solution $ M_{nm} = c_n \delta_{nm}$. The
Hermiticity and positivity constraints restrict finally the
freedom of the choice of the optional sequence of $c_n$ to the
real and positive parameters $c_n \equiv \eta_n> 0$. {\it Vice
versa}, {\em any} choice $\eta_n^{(toy)}$ of the latter sequence
defines an eligible operator of the metric,
 \be
 \eta_{(toy)}=
 \sum_{n=1}^\infty\,|n\rangle\rangle\,\eta_n^{(toy)}
 \,\langle\langle n| = \eta^\dagger_{(toy)} > 0\,.
 \label{atlast}
 \ee
We arrive at a climax and summary of our present message. At any
coupling in the allowed range $\xi< \xi_{1\otimes2} \approx 4.44$
the eigenstates $|n\rangle=|n, \xi\rangle$ and
$|n\rangle\rangle=|n,\xi\rangle\rangle$ pertaining to our toy
model $H=H(\xi)$ have been shown to exhibit an asymptotic
suppression of all the influence of their non-Hermiticity at all
the sufficiently large excitations,
 \be
 |n, \xi\rangle\rangle \approx |n, 0\rangle\rangle,
 \ \ \ \ \
 |n,\xi\rangle \approx |n, 0\rangle\,,
 \ \ \ \ \ \ \ n > N=N(\xi)\,.
 \label{apprp}
 \ee
The value of the cut-off $N(\xi)$ has been shown to grow quite
slowly both with the non-Hermiticity $\xi$ and with the required
precision of the approximation (\ref{apprp}). We may conclude that
on an optional level of precision [and assuming that we choose
$\eta_n^{(toy)}=1$ in eq. (\ref{atlast}) for all $n > N(\xi)$] we
may always treat any pertaining metric $\eta_{(toy)}$ as an
operator which differs from the unit operator just by a
finite-dimensional, $N-$parametric separable modification,
 \be
 \eta_{(toy)} \approx I -
 \sum_{n=1}^{N(\xi)}\,|n,0 \rangle\rangle\,\langle\langle n, 0|+
 \sum_{n=1}^{N(\xi)}\,|n,\xi \rangle\rangle\,\eta_n^{(toy)}
 \,\langle\langle n, \xi|
  \,.
 \label{adslast}
 \ee
This is the property which our toy model shares with the other
non-Hermitian square-well models as available and studied in the
recent literature \cite{sqw,sqwapp,sqwmetric,Langer}. Still, as a
rule, the majority of properties of the latter models looks {\em
much} less accessible to any non-numerical treatment. In contrast
to that, the present energies as well as wave functions preserve a
compact and closed form, particularly transparent in the most
symmetric case with $a=1/2$. For illustrative purposes this
enabled us to construct, entirely systematically and in a fully
non-numerical manner, an explicit sample of the order-by-order
corrections in the approximations (\ref{apprp}). Thus, our model
as well as formulae may be expected to find an immediate
application in the key separable formula (\ref{adslast}) which
acquires, in this manner, the form of an explicit
perturbation-series recipe which may be systematically improved up
to (in principle, arbitrary) precision.

\section*{Acknowledgments}

Participation of M. Z. supported by GA AS (Czech Republic),
grant Nr. A 104 8302.

\end{document}